\documentclass[conference]{IEEEtran}
\AtBeginDocument{%
  }

\usepackage{cite}
\usepackage{amsmath}
\usepackage{booktabs}
\usepackage{graphicx}
\usepackage{url}
\usepackage{tikz}
\usetikzlibrary{positioning,arrows.meta,fit,backgrounds,calc,shapes.geometric}
\usepackage[ruled,linesnumbered,vlined]{algorithm2e}
\SetAlFnt{\small}
\SetAlCapFnt{\small}
\SetAlCapNameFnt{\small}

\begin{document}

\title{Infernux: A Python-Native Game Engine\\with JIT-Accelerated Scripting}
\author{
  \IEEEauthorblockN{Lizhe Chen}
  \IEEEauthorblockA{
    Shenzhen International Graduate School, Tsinghua University\\
    Shenzhen, Guangdong, China\\
    Email: chenlizheme@outlook.com
  }
}

\maketitle

\begin{figure*}[!htbp]
\centering
\includegraphics[width=\textwidth]{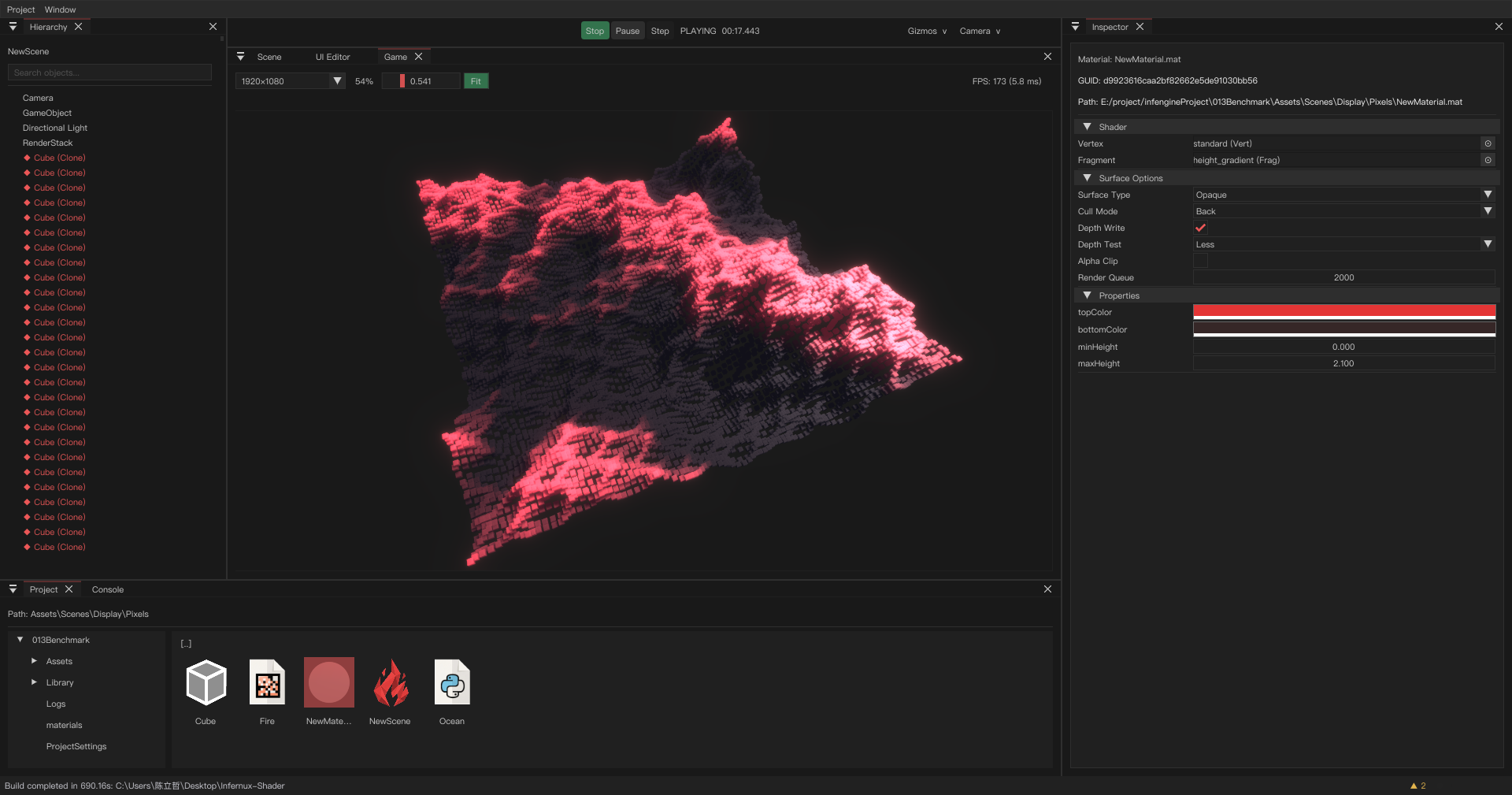}
\caption{The Infernux editor running a 10\,000-cube ocean-FFT demo
  written entirely in Python. All gameplay scripting, editor tooling, and
  render-pipeline definition reside in Python; the C++17/Vulkan core
  handles rendering and physics.}
\label{fig:teaser}
\end{figure*}

\begin{abstract}
This report describes Infernux, an open-source game engine that pairs
a C++17/Vulkan real-time core with a Python production layer connected
through a single pybind11 boundary.  To close the throughput gap
between Python scripting and native-code engines, Infernux combines
two established techniques---batch-oriented data transfer and JIT
compilation---into a cohesive engine-level integration:
(i)~a \emph{batch data bridge} that transfers per-frame state into
contiguous NumPy arrays in one boundary crossing, and (ii)~an optional
\emph{JIT path} via Numba that compiles annotated update functions to
LLVM machine code with automatic loop parallelization.  We compare
against Unity~6 as a reference on three workloads; readers should note
differences in shading complexity, draw-call batching, and editor
tooling maturity between the two engines.  Infernux is MIT-licensed and
available at \url{https://chenlizheme.github.io/Infernux/}.
\end{abstract}


\section{Introduction}

Game engines are now integral to physical
simulation~\cite{Todorov2012MuJoCo}, robot
learning~\cite{Makoviychuk2021IsaacGym}, embodied AI
research~\cite{Juliani2018UnityML}, and real-time visualization, yet
mainstream engines do not treat Python---the dominant language in these
domains~\cite{Paszke2019PyTorch,Harris2020NumPy,OpenAIGym2016}---as a
first-class production language.  Unity~\cite{Unity2024} requires C\#;
Unreal Engine~\cite{UnrealEngine2024} requires C++.  Python can be
connected through IPC bridges or headless simulators, but the engine's
editor, asset pipeline, and rendering remain on the far side of
that boundary, forcing teams to maintain dual codebases.

We present \textsc{Infernux}, an open-source engine designed around
Python as its authoring language.  A single pybind11 boundary separates
a C++17/Vulkan core---rendering, physics, audio, resource
management---from a Python layer that owns gameplay scripting, editor
tooling, render-pipeline definition, and content workflows.  Two
performance mechanisms make this practical at real-time rates: a
\emph{batch data bridge} that transfers engine state into contiguous
NumPy memory in a single crossing per frame, and an optional
\emph{JIT path} via Numba~\cite{Lam2015Numba} that compiles annotated
functions to LLVM machine code with automatic loop parallelization.

\smallskip\noindent\textbf{Key features.}
(1)~A Python-native engine architecture where gameplay scripting,
editor tooling, and render-pipeline definition live in Python while a
C++17/Vulkan core provides the real-time runtime.
(2)~An engine-level integration of batch data transfer and Numba JIT
compilation---with automatic fallback and hot-reload support---that
makes Python scripting practical for per-frame workloads at scale.
(3)~Performance measurements against Unity~6 (IL2CPP) on three
workloads, together with an explicit discussion of confounding factors.

\section{Related Work}\label{sec:related}

\textbf{Game engines.}\quad
Unity~\cite{Unity2024} and Unreal Engine~\cite{UnrealEngine2024}
dominate commercial production but bind developers to C\# and C++,
respectively.  Godot~\cite{Godot2024} offers GDScript, a
Python-inspired language, yet does not load the CPython interpreter or
its package ecosystem.

\textbf{Python simulation environments.}\quad
MuJoCo~\cite{Todorov2012MuJoCo} and NVIDIA Isaac
Gym~\cite{Makoviychuk2021IsaacGym} provide high-throughput physics
backends with Python bindings, but target headless simulation; they
lack interactive editors, asset pipelines, and real-time rendering
loops.  Unity ML-Agents~\cite{Juliani2018UnityML} bridges Unity and
Python via IPC, incurring serialization overhead every step.
EmbodiChain~\cite{EmbodiChain2025} pairs a GPU-accelerated C++ backend
with a Python front-end for embodied AI but likewise omits editing and
general-purpose game development facilities.

\textbf{JIT compilation for Python.}\quad
Numba~\cite{Lam2015Numba} compiles a subset of Python and NumPy to
LLVM machine code, supporting explicit data-parallel loops via its
\emph{prange} construct.  Batch-oriented C++/Python interop through
pybind11 and NumPy is likewise standard practice in scientific
computing~\cite{Harris2020NumPy}.  Our contribution is not the
individual techniques but their integration into a real-time engine
loop: the AST rewriter (Section~\ref{sec:jit}) automates the
\emph{range}$\,\to\,$\emph{prange} promotion so that users need
not annotate parallelism manually.

\textbf{Scripting in games.}\quad
Lua~\cite{Anderson2008Lua} is widely used for embedded game scripting
but lacks the scientific-computing ecosystem of Python.
Domain-specific languages such as GDScript~\cite{GDScript2024} optimise
for engine integration at the cost of library availability.

\section{System Architecture}\label{sec:architecture}

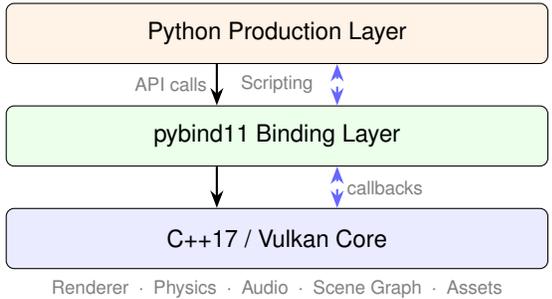
\begin{figure}[!htbp]
\centering
\begin{tikzpicture}[
    >=Stealth,
    layer/.style={draw, rounded corners=3pt, minimum width=7.2cm,
                  minimum height=0.8cm, font=\small\sffamily,
                  align=center, text=black},
    arr/.style={->, thick},
    darr/.style={<->, thick, dashed, blue!60},
    note/.style={font=\scriptsize\sffamily, text=gray},
  ]
  \node[layer, fill=orange!10] (py)
    {Python Production Layer};
  \node[layer, fill=green!8, below=0.55cm of py] (bind)
    {pybind11 Binding Layer};
  \node[layer, fill=blue!8, below=0.55cm of bind] (core)
    {C++17 / Vulkan Core};

  \node[note, below=0.02cm of py.south]
    {Scripting};
  \node[note, below=0.02cm of core.south]
    {Renderer\enspace$\cdot$\enspace Physics\enspace$\cdot$\enspace
     Audio\enspace$\cdot$\enspace Scene Graph\enspace$\cdot$\enspace
     Assets};

  \draw[arr]  ([xshift=-0.8cm]py.south)   -- ([xshift=-0.8cm]bind.north)
    node[midway,left,note] {API calls};
  \draw[arr]  ([xshift=-0.8cm]bind.south) -- ([xshift=-0.8cm]core.north);
  \draw[darr] ([xshift= 0.8cm]core.north) -- ([xshift= 0.8cm]bind.south)
    node[midway,right,note] {callbacks};
  \draw[darr] ([xshift= 0.8cm]bind.north) -- ([xshift= 0.8cm]py.south);
\end{tikzpicture}
\caption{Three-layer architecture.  Solid arrows show the dominant call
  direction (Python\,$\to$\,C++); dashed arrows show engine-to-Python
  callbacks (lifecycle events, physics contacts).  The binding layer is
  the sole crossing point.}
\label{fig:arch}
\end{figure}

Infernux is organized into three layers (Fig.~\ref{fig:arch}).  A
C++17 \emph{native core} owns every latency-sensitive subsystem
(Vulkan rendering, physics, audio, scene graph, assets).  A
\emph{binding layer} (pybind11~\cite{pybind11}) projects these into
the CPython address space.  A \emph{Python production layer}
implements gameplay components, editor panels, render-pipeline
topology, and asset workflows.  The binding layer is the sole crossing
point; C++ subsystems communicate by direct calls, and Python is
shielded from internal native structures.

\subsection{Main Loop}

The native core follows a deterministic initialization order and a
fixed-cadence main loop (Algorithm~\ref{alg:mainloop}).  Physics and
gameplay logic run inside a fixed-timestep accumulator so that
simulation behavior is frame-rate independent.  Rendering and audio
update once per displayed frame.

\begin{algorithm}[!htbp]
\caption{Engine main loop (simplified).}\label{alg:mainloop}
\DontPrintSemicolon
\SetKwProg{Fn}{procedure}{}{}
\SetKwFunction{Loop}{MainLoop}

\Fn{\Loop{}}{
  \While{\normalfont not quit}{
    $\Delta t \gets$ TimeSinceLastFrame()\;
    PollInput()\;
    $\mathit{acc} \mathrel{+}= \Delta t$\;
    \While{$\mathit{acc} \geq \Delta t_{\mathrm{fix}}$}{
      PhysicsStep($\Delta t_{\mathrm{fix}}$)\;
      \ForEach{component $c$}{$c$.\textsc{FixedUpdate}()\;}
      $\mathit{acc} \mathrel{-}= \Delta t_{\mathrm{fix}}$\;
    }
    \ForEach{component $c$}{$c$.\textsc{Update}($\Delta t$)\;}
    CoroutineScheduler.Tick($\Delta t$)\;
    \ForEach{component $c$}{$c$.\textsc{LateUpdate}($\Delta t$)\;}
    CompileRenderGraph()\;
    DrawFrame()\;
  }
}
\end{algorithm}

\subsection{Scene Graph and Component Model}

The scene graph is a forest of game objects managed by a singleton
scene manager that supports multiple loaded scenes and persistent
cross-scene objects.  Components follow a deterministic lifecycle
(\textsc{Awake}\,$\to$\,\textsc{Start}\,$\to$\,per-frame
updates\,$\to$\,\textsc{OnDestroy}).  The engine ships 14 native
component types (transforms, cameras, lights, renderers, rigid
bodies, colliders, audio) plus two bridge components that host
Python-side logic.

\subsection{Python Production Layer}

Gameplay components are Python subclasses of a base component class
assembled from five mixins (native bridge, lifecycle, physics
callbacks, coroutines, serialization).  Serializable fields are
declared through a descriptor mechanism carrying metadata across
17~supported types; the inspector generates matching widgets
automatically.  A coroutine scheduler provides yield-based cooperative
multitasking.

\subsection{Cross-Language Bridge}

Each Python component is mirrored by a C++ proxy that acquires the
GIL, dispatches lifecycle or physics callbacks, and releases the lock.
Component dependencies are resolved at proxy creation; scene state
(hierarchy, native and Python components) is serialized to JSON.

\section{Rendering Pipeline}\label{sec:rendering}

The renderer targets Vulkan~1.3~\cite{VulkanSpec2024} with
VMA~\cite{VMA2024} for GPU memory management and triple-buffered
presentation.  Three design decisions distinguish the pipeline: a
Python-defined topology with \emph{injection points}
(Section~\ref{sec:rendergraph}), a unified post-processing insertion
scheme (Section~\ref{sec:postproc}), and an annotation-driven shader
composition system (Section~\ref{sec:shader}).

\subsection{Render Graph with Injection Points}\label{sec:rendergraph}

Frame composition follows a declarative graph inspired by Frostbite's
FrameGraph~\cite{OConnor2017Vulkan}.  The key departure is that the
graph is \emph{split-authored}: a Python pipeline subclass defines the
\emph{topology} (pass ordering, resource declarations, action types), while
a C++~compiler handles DAG optimization, dead-pass culling, barrier
insertion, and transient-resource aliasing.  This split lets users
redefine the rendering strategy without recompiling the engine.

\paragraph{Pass definition.}
Each pass declares an action type (e.g.\ draw renderers, fullscreen
quad, compute), resource reads/writes, queue-range filters, and sort
mode.  Multiple colour targets per pass enable G-buffer layouts.

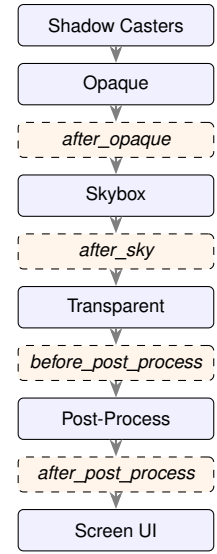
\begin{figure}[!htbp]
\centering
\begin{tikzpicture}[
    >=Stealth,
    pass/.style={draw, rounded corners=2pt, fill=blue!6,
                 minimum width=2.6cm, minimum height=0.55cm,
                 font=\scriptsize\sffamily, align=center},
    ip/.style={draw, dashed, rounded corners=2pt, fill=orange!8,
               minimum width=2.6cm, minimum height=0.45cm,
               font=\scriptsize\sffamily\itshape, align=center},
    arr/.style={->, thick, gray},
  ]
  \node[pass]                      (shd) {Shadow Casters};
  \node[pass, below=0.22cm of shd] (opq) {Opaque};
  \node[ip,   below=0.22cm of opq] (ip1) {after\_opaque};
  \node[pass, below=0.22cm of ip1] (sky) {Skybox};
  \node[ip,   below=0.22cm of sky] (ip2) {after\_sky};
  \node[pass, below=0.22cm of ip2] (trn) {Transparent};
  \node[ip,   below=0.22cm of trn] (ip3) {before\_post\_process};
  \node[pass, below=0.22cm of ip3] (pp)  {Post-Process};
  \node[ip,   below=0.22cm of pp]  (ip4) {after\_post\_process};
  \node[pass, below=0.22cm of ip4] (ui)  {Screen UI};

  \foreach \a/\b in {shd/opq, opq/ip1, ip1/sky, sky/ip2,
                      ip2/trn, trn/ip3, ip3/pp, pp/ip4, ip4/ui}
    \draw[arr] (\a) -- (\b);
\end{tikzpicture}
\caption{Forward-pipeline topology.  Solid boxes are built-in passes;
  dashed boxes are injection points.  User-mounted effects or custom
  passes attach at any injection point via the resource bus.}
\label{fig:forward_topo}
\end{figure}

\paragraph{Injection points.}
An \emph{injection point} is a named slot where additional passes may
be inserted without modifying the pipeline subclass.  The author
declares each slot with a \emph{resource contract}---textures
guaranteed available in the \emph{resource bus}.  Two default points
(\emph{before/after\_post\_process}) are auto-injected if not declared
explicitly.  Fig.~\ref{fig:forward_topo} shows the forward topology
with its four injection points.

\paragraph{Resource bus.}
A \emph{resource bus} carries texture handles between passes.
Effects read from the bus, modify data, and write back; subsequent
effects automatically receive updated handles, eliminating explicit
inter-effect dependency declarations.

\paragraph{DAG compilation.}
The recorded graph is serialized and handed to a C++ compiler
(Algorithm~\ref{alg:graphcompile}).  Phase~1 performs a standard
backward-reachability analysis: the $\mathit{live}$ set is
initialized to the final output resource and grown by the reads of
each reachable pass.  A pass is reachable if and only if at least
one of its declared write targets is currently in $\mathit{live}$;
this ensures that passes whose outputs are never consumed---directly
or transitively---are culled.  The reverse traversal order guarantees
that when a pass is visited, all passes that could consume its
outputs have already been evaluated.

Phase~3 inserts Vulkan pipeline barriers between dependent passes.
Barrier types are determined by resource-usage transitions:
a colour-attachment write followed by a shader read yields
appropriate source/destination stage masks and a memory barrier
ensuring visibility.  Compute-to-compute transitions use the
analogous compute-shader stage bits.  Where possible, the compiler
merges adjacent barriers and promotes image layout transitions to
a general layout to avoid redundant transitions within a single
render pass.

\begin{algorithm}[!htbp]
\caption{Render-graph compilation.}\label{alg:graphcompile}
\DontPrintSemicolon
\SetKwProg{Fn}{procedure}{}{}
\SetKwFunction{Compile}{CompileGraph}

\Fn{\Compile{passes $P$, output $O$}}{
  \tcp{Phase 1: backward cull}
  $\mathit{live} \gets \{O\}$\;
  \ForEach{pass $p$ in reverse order}{
    \If{$p$ writes a resource in $\mathit{live}$}{
      mark $p$ as reachable\;
      $\mathit{live} \gets \mathit{live} \cup \mathrm{reads}(p)$\;
    }
  }
  remove unreachable passes\;
  \BlankLine
  \tcp{Phase 2: resource lifetimes}
  \ForEach{resource $r$}{
    record first-pass and last-pass usage\;
  }
  \BlankLine
  \tcp{Phase 3: topological sort (Kahn)}
  build edges: writer($r$) $\to$ reader($r$)\;
  sort passes by in-degree; insert barriers\;
  \Return{sorted pass list with barriers}\;
}
\end{algorithm}

\paragraph{Built-in presets.}
A \emph{forward} pipeline (Fig.~\ref{fig:forward_topo}) with optional
MSAA and a \emph{deferred} pipeline with a four-target G-buffer are
provided.  Both share the same injection-point names, so user effects
are portable across presets.  Directional shadows use four-cascade
CSM~\cite{Engel2006CSM}.

\subsection{Post-Processing via Injection Points}\label{sec:postproc}

Each post-processing effect is an independent unit that attaches to an
injection point and interacts with the resource bus, decoupling effect
authoring from pipeline authoring.  Every effect declares three
contracts---\emph{requires}, \emph{modifies}, and an injection-point
name---and is sorted by priority at graph-build time.

Complex effects may inject multiple passes; bloom, for example,
generates eight passes (prefilter, Jimenez
downsamples~\cite{Jimenez2014Bloom}, tent upsamples, composite) with
transient intermediate textures whose lifetimes are managed by the
graph compiler.  Eight effects ship with the engine (bloom, tone
mapping, chromatic aberration, colour adjustments, film grain, sharpen,
vignette, white balance).

\subsection{Annotation-Driven Shader Composition}\label{sec:shader}

Modern shader systems balance composability with
transparency.  Slang~\cite{glslang2024} uses runtime reflection;
Infernux takes a different approach: \emph{compile-time source
rewriting} driven by in-source annotations.  The loader recognizes
15~directives (shader identity, shading model, surface options,
material properties, imports) prefixed with~\emph{@}; annotations are
stripped before GLSL compilation.

\begin{figure}[!htbp]
\centering
\begin{tikzpicture}[
    >=Stealth,
    stage/.style={draw, rounded corners=3pt, fill=blue!6,
                  minimum width=5.8cm, minimum height=0.65cm,
                  font=\small\sffamily, align=center},
    arr/.style={->, thick},
    note/.style={font=\scriptsize\sffamily, text=gray, align=left},
  ]
  \node[stage] (s1) {1. Parse annotations $\to$ descriptor IR};
  \node[stage, below=0.6cm of s1] (s2)
    {2. Resolve imports (recursive, dedup)};
  \node[stage, below=0.6cm of s2] (s3)
    {3. Generate target-specific GLSL};

  \draw[arr] (s1) -- (s2);
  \draw[arr] (s2) -- (s3);

  \node[note, right=0.15cm of s1.east]
    {shader\_id, properties,\\surface options, imports};
  \node[note, right=0.15cm of s2.east]
    {diamond dedup, depth$\,\leq\,$16,\\version-directive stripping};
  \node[note, right=0.15cm of s3.east]
    {forward / G-buffer / shadow\\variants from one source};
\end{tikzpicture}
\caption{Three-stage shader preprocessing pipeline.  Annotations are
  parsed into a structured descriptor, imports are recursively inlined
  with diamond deduplication, and target-specific GLSL is emitted for
  each compilation variant.}
\label{fig:shader_pipeline}
\end{figure}
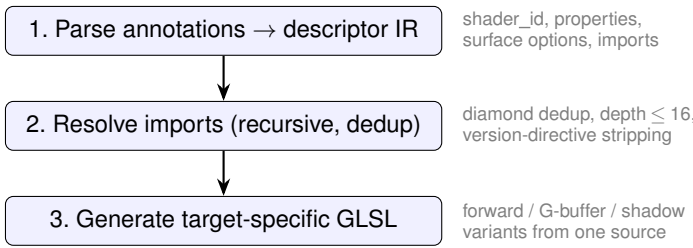

\paragraph{Import resolution.}
The \emph{@import} directive inlines the file matching a referenced
shader identifier (Fig.~\ref{fig:shader_pipeline}).

Resolution uses a \emph{shader-id map} built by scanning project and
engine shader directories; a diamond-deduplication set ensures
transitive imports are inlined exactly once.  The recursion depth is
capped at~16---sufficient for the deepest practical shader graphs we
have encountered (the most complex shipped shader reaches depth~8)---and
cyclic imports are detected by the deduplication set: if a shader
identifier appears twice on the current import stack, the loader emits
a compile-time error with the full cycle trace rather than silently
discarding the import.

\paragraph{Shading-model composition.}
A \emph{shading\_model} annotation selects a model (e.g.\ PBR,
unlit).  Model files contain labelled GLSL fragments for each
compilation target.  The code generator expands a surface shader into
three complete GLSL programs---forward, G-buffer, shadow---with
auto-generated UBO declarations, builtins, and texture samplers.
Unlike reflection-based systems (Slang, SPIRV-Cross) that discover
bindings \emph{after} compilation, source rewriting resolves all UBO
fields, samplers, and push-constant ranges \emph{before} glslang is
invoked, yielding human-readable GLSL and eliminating external
material-state files.

\begin{figure}[!htbp]
\centering
\includegraphics[width=\columnwidth]{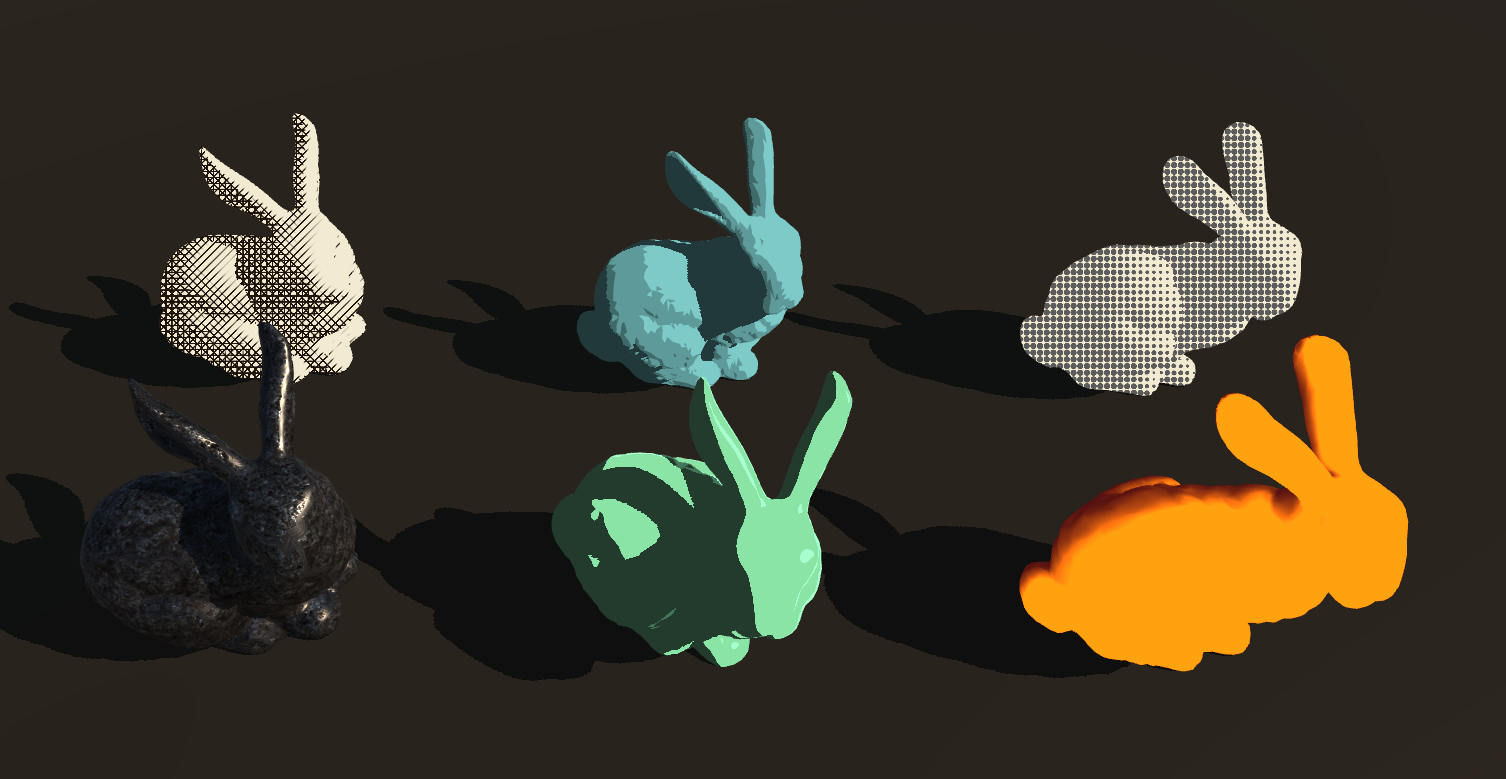}
\caption{Stylized cel-shading rendered with a custom surface shader
  authored through the annotation-driven composition system.  The
  shader uses the \emph{@shading\_model} directive to select an
  unlit toon model, applies a stepped diffuse ramp with rim lighting,
  and adds outline passes via an injection point---all defined in a
  single annotated GLSL source file.}
\label{fig:toon}
\end{figure}

\section{Physics Integration}\label{sec:physics}

Rigid-body dynamics are provided by Jolt Physics~\cite{JoltPhysics2024}.
The engine wraps the Jolt broadphase, a thread-pool job scheduler, and a
256\,MB temporary allocator, supporting up to $2^{16}$ bodies, body
pairs, and contact constraints.  Four collider primitives are provided
(box, sphere, capsule, triangle mesh); compound shapes are supported.
The rigid-body component exposes force application modes (force,
acceleration, impulse, velocity change), per-axis freeze constraints,
and four collision-detection modes including continuous-speculative.
Spatial queries---raycasts, overlap tests, and shape casts---are
exposed through the binding layer with a 32-layer filtering system.

The physics step runs inside the fixed-timestep accumulator
(Algorithm~\ref{alg:mainloop}), decoupled from the display rate.
Collision and trigger callbacks are dispatched to both native and
Python components through the contact listener.

\begin{figure}[!htbp]
\centering
\includegraphics[width=\columnwidth]{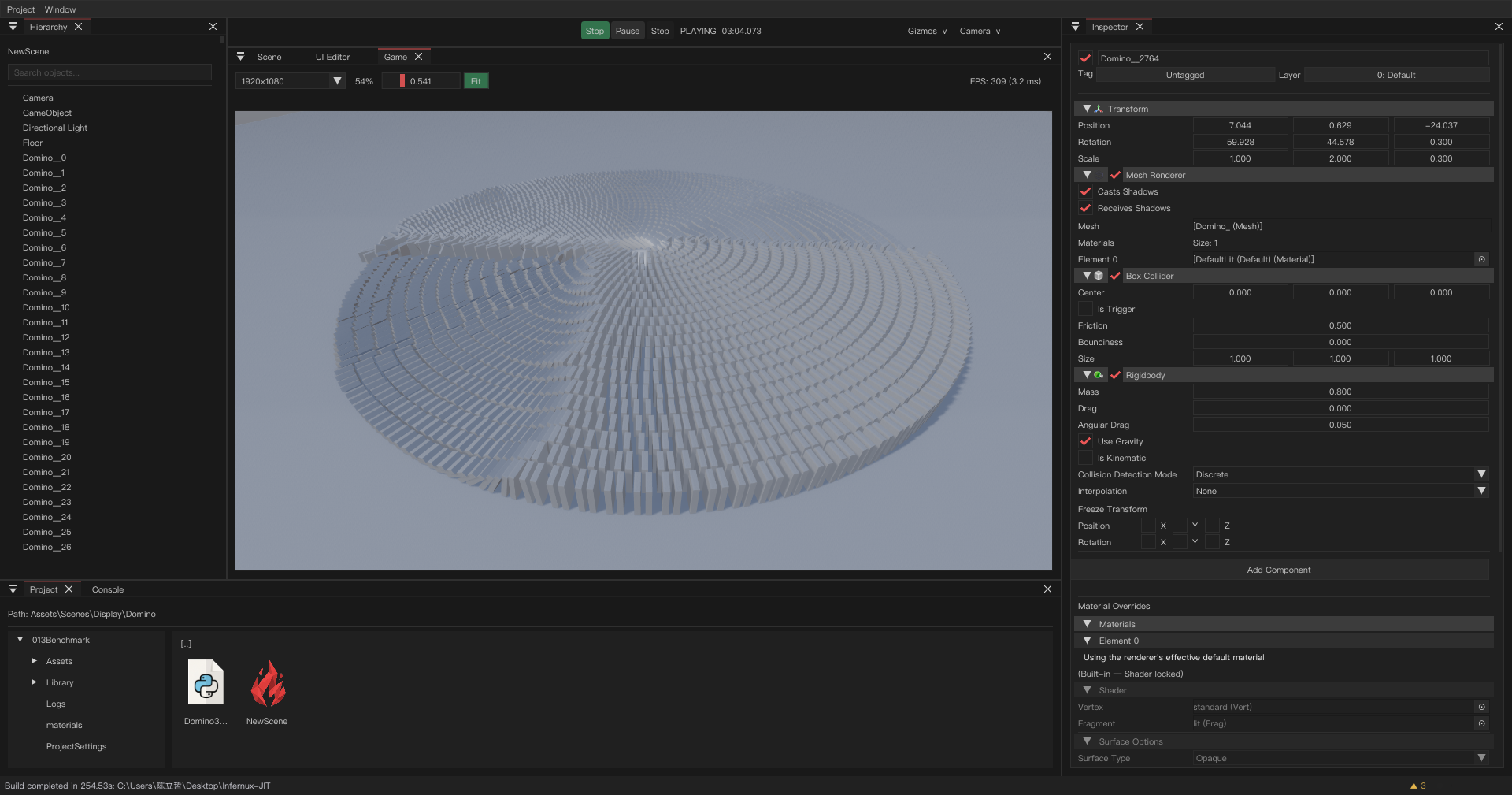}
\caption{A domino-chain simulation driven by Jolt Physics through the
  Infernux binding layer.  Over 500 rigid bodies with box colliders
  topple in sequence; the fixed-timestep accumulator and
  continuous-speculative collision detection maintain stable contacts
  throughout the cascade.}
\label{fig:domino}
\end{figure}

\section{Performance Bridge}\label{sec:bridge}

Pure Python is too slow for per-object, per-frame logic at scale:
each binding call acquires the GIL, converts types, and releases the
lock---overhead that dominates at high object counts.  Infernux
provides two mechanisms: a \emph{batch data bridge} that amortizes
crossing cost across $N$ objects in one call
(Section~\ref{sec:batch}), and an optional \emph{JIT path} that
compiles functions to LLVM machine code with automatic
parallelization (Section~\ref{sec:jit}).

\subsection{Batch Data Bridge}\label{sec:batch}

The batch API collapses $N$ individual property accesses into one
round-trip across the language boundary.  A read or write selects the
fastest available path through a three-tier dispatch:

\begin{enumerate}
  \item \textbf{Native transform path.}\quad
    Reads/writes go to a C++ SoA store indexed by generational
    handles; the GIL is released during the native gather/scatter
    kernel.  Thread safety is guaranteed by the handle mechanism:
    if Python-side code destroys a game object while a gather is in
    progress, the generational tag mismatches and the slot is
    skipped, avoiding use-after-free.
  \item \textbf{Component data store.}\quad
    Numeric fields use a per-class C++ SoA with $O(1)$ slot access
    and six typed lanes (32- and 64-bit floats, 32- and 64-bit
    integers, booleans, and a packed four-float lane for vectors
    and quaternions).
  \item \textbf{Interpreted fallback.}\quad
    Per-element Python attribute access via the standard pybind11
    property path; used only for non-numeric or user-defined fields.
\end{enumerate}

The batch API exposes two calls---\emph{batch\_read} and
\emph{batch\_write}---each taking a handle array and a field-name
string.  The field name is resolved to a column index through a
per-class hash map built once at component registration; subsequent
frames reuse the cached index, so the per-call lookup cost is~$O(1)$.
Returned arrays use 32-bit floats with shapes $(N{,}k)$ for
$k$-component vectors.  A handle object caches native pointers to
avoid repeated $O(N)$ binding casts within a single frame.

\subsection{JIT Compilation and Auto-Parallelization}\label{sec:jit}

The engine provides a decorator that wraps
Numba~\cite{Lam2015Numba} with three additions:
\emph{(i)}~graceful degradation to pure Python when Numba is absent
(a runtime warning is emitted on first invocation to alert the user
to the performance impact);
\emph{(ii)}~a bytecode-keyed compilation cache that survives module
reloads and hot-reload cycles; and
\emph{(iii)}~partial compatibility with Nuitka~\cite{Nuitka2024}
ahead-of-time compilation.  In the Nuitka case the original
source file may not exist on disk, so the decorator falls
back to bytecode-based cache keys; the Numba JIT itself still
requires the CPython interpreter at runtime, so Nuitka serves
primarily as a distribution and startup-time optimization rather than
a full AOT replacement for the JIT path.

\paragraph{Auto-parallel mode.}
When the auto-parallel flag is set, an AST rewriter examines each
\emph{counted loop}---defined as a ``\verb|for i in range(n)|''
construct with an integer induction variable.  Iterator-based loops
(``\verb|for x in arr|'') and \verb|while| loops are not
rewritten; these forms lack the fixed trip count that Numba's
\emph{prange} requires.  If the loop body contains only indexed
array stores and supported reductions (sum, product, min, max), and
no unsupported control flow (early return, \verb|break|,
\verb|yield|, exception handling), the rewriter promotes
\emph{range} to \emph{prange} and enables Numba's thread-level
parallelism.

The ``disjoint memory locations'' requirement deserves clarification.
The AST rewriter performs a \emph{syntactic} check: it verifies that
all store targets use the loop induction variable as the leading
index and that no two stores target the same array-and-index
expression.  This is \emph{not} a full alias analysis; it cannot
detect aliasing through indirection or non-trivial index arithmetic.
At runtime, Numba's parallel semantics assume that \emph{prange}
iterations write to non-overlapping memory (analogous to the OpenMP
shared-memory model); if this assumption is violated, results are
undefined.  We consider this an acceptable trade-off because the
dominant use case---writing to position, rotation, or scale arrays
indexed by entity ID---naturally satisfies the non-aliasing
constraint.

Both a serial and a parallel variant are compiled; the runtime
defaults to the parallel path and falls back to the serial variant
on error.  Compiling two variants increases cold-start latency
(typically 50--200\,ms per decorated function on our test hardware),
but a warm-up helper that pre-compiles all registered functions
during scene load eliminates this cost from the interactive loop.
In hot-reload scenarios, only the modified function is recompiled;
unchanged functions retain their cached machine code.

\subsection{Combined Workflow}

The typical workflow reads transforms into a NumPy array via one batch
call, passes it to a JIT kernel (GIL released), and writes back via
one batch call---two boundary crossings per frame regardless of object
count (Algorithm~\ref{alg:batchjit}).

\begin{algorithm}[!htbp]
\caption{Batch\,+\,JIT per-frame update.}\label{alg:batchjit}
\DontPrintSemicolon
\SetKwProg{Fn}{procedure}{}{}
\SetKwFunction{Up}{Update}

\Fn{\Up{targets $T$, time $t$}}{
  $\mathbf{p} \gets$ BatchRead($T$, position) \tcp*{$(N,3)$ array}
  JIT\_WaveKernel($\mathbf{p}$, $t$, $N$) \tcp*{GIL released}
  BatchWrite($T$, $\mathbf{p}$, position)\;
}
\end{algorithm}

\section{Evaluation}\label{sec:evaluation}

We evaluate Infernux on three benchmarks and compare against Unity~6.

\subsection{Environment}

All measurements are collected on a single workstation:
Intel Core Ultra~7 265K (24 cores), NVIDIA GeForce RTX~5070\,Ti,
64\,GB DDR5-6400, Windows~11 (Build~26200), Vulkan~1.3.
The Infernux build uses CPython~3.12 with Numba~0.60 (LLVM~15);
the Unity build uses Unity~6 (6000.1) with the IL2CPP scripting
backend.  V-Sync is disabled; resolution is $1920 \times 1080$.
All tables report mean FPS over 300~frames after a 60-frame warm-up.

\subsection{Experiment~1: SpawnCube (Single Material)}

An $N \times N$ grid of cubes over a $50 \times 50$\,m area is driven
by a dual-axis sinusoidal-wave kernel every frame:
\begin{equation}
  y_i = A\sin(\omega\, x_i + v\, t) + \tfrac{A}{2}\sin(\omega\, z_i + 1.3\,v\, t),
\end{equation}
where $A$ is the amplitude, $\omega$ the spatial frequency, $v$ the
wave speed, and $(x_i, z_i)$ the base grid position of cube~$i$.
All cubes share one material and one directional light, minimizing
draw-call overhead so that per-frame scripting cost is the primary
variable---though differences in shading complexity between the two
engines remain a confounding factor (see Analysis).  The Infernux side
uses the batch\,+\,JIT path (Section~\ref{sec:bridge}); Unity uses an
equivalent C\# MonoBehaviour.  $N$ ranges from~10 to~100.

\begin{table}[!htbp]
  \caption{Experiment~1: SpawnCube, single material.  Higher is better.}
  \label{tab:fps}
  \centering\small
  \begin{tabular*}{\columnwidth}{@{\extracolsep{\fill}}rcccc@{}}
    \toprule
    & \multicolumn{2}{c}{Infernux (Batch+JIT)}
    & \multicolumn{2}{c}{Unity~6 (IL2CPP)} \\
    \cmidrule(lr){2-3}\cmidrule(lr){4-5}
    $N{\times}N$ & Editor & Runtime & Editor & Runtime \\
    \midrule
    $10^2$   & 803  & $>$1000 & 714  & $>$1000 \\
    $30^2$   & 625  & 865   & 404  & $>$1000 \\
    $50^2$   & 414  & 647   & 228  & 651 \\
    $70^2$   & 265  & 437   & 125  & 347 \\
    $100^2$  & 127  & 171   & 61   & 187 \\
    \bottomrule
  \end{tabular*}
\end{table}

\subsection{Experiment~2: SpawnCube (Multi-Material Variant)}

This experiment uses the same wave kernel as Experiment~1 but assigns
$M$~distinct PBR materials (differing in base colour) in a round-robin
pattern, forcing pipeline-state switches and draw-call splitting.
The goal is to measure how \emph{rendering-pipeline} overhead scales
with material count while scripting cost remains constant.  Because
Infernux lacks the draw-call batching present in Unity's SRP Batcher,
results primarily reflect each engine's draw-dispatch strategy rather
than scripting throughput alone.

\begin{table}[!htbp]
  \caption{Experiment~2: SpawnCube with $M{=}10$ materials (FPS).}
  \label{tab:fps_mat10}
  \centering\small
  \begin{tabular*}{\columnwidth}{@{\extracolsep{\fill}}rcccc@{}}
    \toprule
    & \multicolumn{2}{c}{Infernux}
    & \multicolumn{2}{c}{Unity~6} \\
    \cmidrule(lr){2-3}\cmidrule(lr){4-5}
    $N{\times}N$ & Editor & Runtime & Editor & Runtime \\
    \midrule
    $10^2$   & 778  & 968   & 711  & $>$1000 \\
    $30^2$   & 588  & 812   & 346  & $>$1000 \\
    $50^2$   & 389  & 591   & 194  & 651 \\
    $70^2$   & 229  & 397   & 107  & 339 \\
    $100^2$  & 114  & 162   & 65   & 182 \\
    \bottomrule
  \end{tabular*}
\end{table}

\begin{table}[!htbp]
  \caption{Experiment~2: SpawnCube with $M{=}100$ materials (FPS).}
  \label{tab:fps_mat100}
  \centering\small
  \begin{tabular*}{\columnwidth}{@{\extracolsep{\fill}}rcccc@{}}
    \toprule
    & \multicolumn{2}{c}{Infernux}
    & \multicolumn{2}{c}{Unity~6} \\
    \cmidrule(lr){2-3}\cmidrule(lr){4-5}
    $N{\times}N$ & Editor & Runtime & Editor & Runtime \\
    \midrule
    $10^2$   & 572  & 723   & 678  & $>$1000 \\
    $30^2$   & 458  & 631   & 436  & $>$1000 \\
    $50^2$   & 331  & 484   & 227  & 623 \\
    $70^2$   & 201  & 324   & 119  & 333 \\
    $100^2$  & 109  & 152   & 40   & 178 \\
    \bottomrule
  \end{tabular*}
\end{table}

\subsection{Experiment~3: Pure Computation (No Transform Writes)}

To isolate scripting cost from GPU work, the sin-wave kernel runs on
an in-memory NumPy array without instantiating scene objects or
writing positions back to Transforms.  Grid side length~$N$ scales
from~100 to~1\,000 ($N^{2}{=}10\text{k}$--$1\text{M}$ elements).
We compare the Numba JIT auto-parallel path against the plain NumPy
vectorised path (``No~JIT'') in Infernux, and against Unity's
IL2CPP-compiled C\#~loop.

\begin{table}[!htbp]
  \caption{Experiment~3: pure-compute FPS (no Transform writes).
    $N$ is the grid side length; element count is~$N^{2}$.}
  \label{tab:compute}
  \centering\scriptsize
  \begin{tabular*}{\columnwidth}{@{\extracolsep{\fill}}rcccccc@{}}
    \toprule
    & \multicolumn{4}{c}{Infernux} & \multicolumn{2}{c}{Unity~6} \\
    \cmidrule(lr){2-5}\cmidrule(lr){6-7}
    & \multicolumn{2}{c}{Editor} & \multicolumn{2}{c}{Runtime}
    & Ed. & Rt. \\
    \cmidrule(lr){2-3}\cmidrule(lr){4-5}
    $N$ & JIT & No JIT & JIT & No JIT & IL2CPP & IL2CPP \\
    \midrule
    100  & 766  & 765  & $>$1k & $>$1k & 557  & $>$1k \\
    300  & 755  & 636  & $>$1k & 838   & 172  & 994   \\
    500  & 724  & 472  & 989   & 583   & 72   & 422   \\
    700  & 694  & 143  & 949   & 159   & 38   & 241   \\
    1000 & 624  & 80   & 848   & 81    & 19   & 123   \\
    \bottomrule
  \end{tabular*}
\end{table}

\subsection{Analysis}

\paragraph{Experiment~1 (single material).}
At low density ($N{\leq}30$, $\leq$900 cubes) Unity's IL2CPP runtime
exceeds 1\,000\,FPS because the per-object C\# loop is negligible and
Unity employs a simplified Cook-Torrance BRDF~\cite{Unity2024} that
leaves substantial GPU headroom.  Infernux uses a full-precision PBR
shading model and already pays the fixed pybind11 batch-dispatch cost,
so its runtime starts at $>$1\,000\,FPS only at $N{=}10$.
At medium density ($N{=}50$) the runtime curves converge as both
engines become GPU-bound---both report $\approx$650\,FPS.
At high density ($N{\geq}70$) Unity's runtime retains a
${\sim}9\%$ advantage (187 vs.\ 171\,FPS at $N{=}100$),
consistent with its lighter per-fragment shading cost.

The editor results diverge sharply.  At $N{=}100$ the Infernux editor
sustains 127\,FPS---more than $2{\times}$ the Unity editor's 61\,FPS.
Unity's editor carries a richer tooling layer (scene-view gizmos,
profiler, real-time asset previews, domain reload) that the current
Infernux editor does not yet replicate.  The batch\,+\,JIT path keeps
the Infernux editor above interactive rates at all tested scales.

\paragraph{Experiment~2 (multi-material).}
Increasing the material count from~1 to~10 reduces Infernux runtime
FPS at $N{=}100$ from 171 to 162 ($-5\%$), while Unity drops from
187 to 182 ($-3\%$).  At $M{=}100$ Infernux falls to 152 ($-11\%$
vs.\ baseline) and Unity to 178 ($-5\%$).  Unity's SRP Batcher
minimizes pipeline-state switching by coalescing draw calls that share
the same shader variant, making it nearly insensitive to material
count.  Infernux groups draw calls by material slot and batches within
each group, so additional materials add moderate overhead from
pipeline-bind and descriptor-set switches.  In the editor the gap is
reversed: at $M{=}100$, $N{=}100$ Infernux still sustains 109\,FPS
while Unity's editor falls to 40\,FPS, again because Unity's heavier
editor instrumentation amplifies per-draw overhead.  As noted above,
the editor columns should be interpreted with caution given the
asymmetry in editor tooling maturity.

\paragraph{Experiment~3 (pure compute).}
With no GPU work and no Transform writes, this experiment isolates
scripting throughput.  The Numba JIT auto-parallel path degrades
gracefully: from 766 to 624\,FPS in the editor ($-19\%$) and from
$>$1\,000 to 848\,FPS in the runtime as element count grows from
10\,k to~1\,M\@.  Automatic loop parallelization distributes the
sin-wave kernel across all 24 hardware threads.
The plain NumPy path (``No~JIT'') is single-threaded;
it collapses between $N{=}500$ and $N{=}700$, dropping from 583 to
159\,FPS in a narrow band.  At $N{=}1000$ the JIT runtime reaches
848\,FPS, $6.9{\times}$ faster than Unity's IL2CPP runtime
(123\,FPS) and $10.5{\times}$ the No-JIT path (81\,FPS).
In the editor, Unity drops to 19\,FPS at $N{=}1000$---$33{\times}$
slower than the JIT editor path (624\,FPS).

\paragraph{Cross-language communication overhead.}
Across all three experiments, profiling reveals that the single largest
residual cost is the \emph{Python$\,\leftrightarrow\,$C++
communication boundary} itself.  Every batch dispatch crosses pybind11:
the caller acquires the GIL, marshals SoA column pointers through
pybind11's type-conversion layer, invokes the native batch kernel, and
then releases the GIL on return.  Although individual crossing costs
are small (${\sim}2$--$5\,\mu$s per call on our test hardware), a
frame that issues dozens of batch calls accumulates measurable
overhead.  At $N{=}100$ in Experiment~1, communication accounts for an
estimated 8--12\% of per-frame CPU time; at higher densities the
fraction shrinks as the native kernel work dominates, but the
\emph{absolute} crossing cost remains constant and sets an asymptotic
ceiling on achievable frame rate.

This observation motivates a planned \emph{lock-free binding} path.
The idea is to replace the current GIL-synchronized pybind11 dispatch
with a lock-free command ring: the Python thread enqueues batch
descriptors into a wait-free SPSC (single-producer, single-consumer)
ring buffer, while the native worker thread dequeues and executes them
without ever touching the GIL\@.  Preliminary micro-benchmarks of the
ring-buffer primitive show $<$100\,ns per enqueue--dequeue pair,
suggesting that the communication ceiling can be lowered by at least an
order of magnitude.  We plan to integrate and evaluate this path in a
future release.

\paragraph{Threats to validity.}
Several factors limit the generalizability of these results.
First, the rendering pipelines differ: Infernux uses a full-precision
PBR shading model while Unity uses a simplified Cook-Torrance BRDF;
Experiments~1--2 therefore conflate scripting and rendering costs.
Experiment~3 eliminates GPU work entirely, providing the cleanest
comparison of scripting throughput.
Second, the editor columns are not directly comparable because
Unity's editor provides substantially richer tooling (profiler,
scene-view gizmos, domain reload) whose overhead is absent in the
current Infernux editor; we include editor results for transparency
but recommend focusing on the Runtime columns for cross-engine
comparison.
Third, all measurements are collected on a single hardware
configuration; results may vary on different CPU core counts or GPU
tiers.
Fourth, Unity's SRP Batcher is a mature draw-call optimization that
has no counterpart in Infernux; Experiment~2's multi-material
results therefore measure draw-dispatch strategy as much as they
measure scripting cost.
\section{Discussion}\label{sec:discussion}

\paragraph{Communication boundary as the performance ceiling.}
Our benchmarks consistently show that the dominant remaining overhead
in Infernux is the Python$\,\leftrightarrow\,$C++ communication
boundary rather than either language's arithmetic throughput.
Each pybind11 batch call acquires the GIL, performs type conversion on
SoA column pointers, dispatches the native kernel, and releases the
GIL\@.  While individual calls are inexpensive (${\sim}2$--$5\,\mu$s),
a frame that issues many batch operations accumulates non-trivial
crossing cost.  This cost is \emph{fixed} per call regardless of
element count, so it becomes the asymptotic bottleneck as native
kernel work shrinks---precisely the regime targeted by multi-core JIT
parallelization.  In practice, this means the JIT path cannot fully
amortize its speedup at very high call frequencies, establishing a
ceiling that currently sits at ${\sim}850$--$1\,000$\,FPS for our
heaviest compute workloads.

\paragraph{Toward a lock-free binding layer.}
To push past this ceiling we are prototyping a \emph{lock-free command
ring} that decouples the Python submission thread from the native
execution thread.  The Python side enqueues lightweight batch
descriptors into a wait-free SPSC ring buffer; the C++ worker dequeues
and executes them without acquiring the GIL at any point.  Preliminary
micro-benchmarks of the ring primitive measure $<$100\,ns per
enqueue--dequeue pair, roughly $20$--$50{\times}$ cheaper than the
current pybind11 round-trip.  Integrating this path requires a
double-buffered SoA store so that the Python thread can prepare
frame~$n{+}1$ while the native thread consumes frame~$n$;
the generational-handle scheme already in place
(Section~\ref{sec:bridge}) provides the necessary aliasing safety.

\paragraph{Draw-call batching.}
Experiment~2 shows that Infernux's per-material draw
dispatch still incurs moderate overhead ($-11\%$ at $M{=}100$),
whereas Unity's SRP Batcher coalesces draw calls that share the same
shader variant into a single GPU submission and is nearly unaffected.

\paragraph{GIL considerations.}
The engine currently releases the GIL during all native work;
the Numba JIT path likewise executes compiled machine code in
GIL-released mode.  Python~3.13 introduces an experimental
free-threaded build (PEP~703) that removes the GIL entirely.
If this build stabilizes, it would eliminate the GIL-acquisition
cost in the pybind11 path and allow true concurrent Python threads
to feed the batch bridge from multiple cores---a scenario we intend
to evaluate once the free-threaded ABI matures.

\paragraph{CPython runtime costs.}
Beyond the GIL, the CPython runtime imposes additional costs that
affect a real-time engine.  Reference counting triggers atomic
increments and decrements on every temporary object, polluting CPU
caches with write traffic to widely scattered reference-count
fields~\cite{Harris2020NumPy}.  The cyclic garbage collector, while
infrequent, can cause unpredictable frame-time spikes when it scans
large object graphs.  To mitigate these costs, Infernux disables the
cyclic GC during the frame loop and triggers collection explicitly
between scenes; long-lived engine objects (transforms, meshes,
materials) are stored entirely on the C++ side and are invisible to
the Python GC\@.  Memory fragmentation from CPython's small-object
allocator (\emph{pymalloc}) is a further concern for long-running
sessions; we have not yet observed pathological fragmentation in
practice but plan to investigate arena-based allocation for component
data in future work.

\paragraph{Platform scope and missing subsystems.}
The current release targets Windows with Vulkan~1.3.  macOS support
via MoltenVK and Linux support via native Vulkan on SDL3 are under
active development and share $>$95\% of the rendering back-end code.
Skeletal animation, GPU particle systems, spatial audio, and
networking are planned for future milestones; their absence currently
limits the engine's applicability to shipping titles but does not
affect the measurements presented here.

\subsection{Roadmap}

The following items are under active development or planned for near-term
releases.

\begin{enumerate}
  \item \textbf{Lock-free command ring.}\quad
    Replace the GIL-synchronized pybind11 dispatch with a wait-free
    SPSC ring buffer, targeting $<$100\,ns per dispatch (see
    Section~\ref{sec:discussion}).
  \item \textbf{Material-slot draw-call batcher.}\quad
    Group meshes by pipeline state and emit indirect draw calls,
    closing the gap with Unity's SRP Batcher exposed by Experiment~2.
  \item \textbf{Free-threaded Python (PEP~703).}\quad
    Evaluate the GIL-free CPython build once the ABI stabilizes.
  \item \textbf{Arena allocator for component data.}\quad
    Mitigate \emph{pymalloc} fragmentation in long-running sessions.
  \item \textbf{Cross-platform support.}\quad
    macOS (MoltenVK) and Linux (native Vulkan / SDL3).
  \item \textbf{Missing subsystems.}\quad
    Skeletal animation, GPU particle systems, spatial audio,
    networking.
  \item \textbf{ML integration.}\quad
    PyTorch--Vulkan tensor interop, GPU compute dispatch from the
    Python render graph, and a Gym-compatible environment wrapper.
\end{enumerate}

\section{Conclusion}\label{sec:conclusion}

This report has described Infernux, an open-source game engine whose
architecture places Python at the center of the authoring workflow
while delegating latency-sensitive subsystems---rendering, physics,
and asset I/O---to a C++17/Vulkan core.  Three design features make
this combination practical for real-time workloads.

First, a \emph{batch data bridge} with SoA column stores and
generational handles amortizes the Python$\,\leftrightarrow\,$C++
crossing cost by transferring thousands of entity attributes in a
single pybind11 call, avoiding the per-object overhead that plagues
na\"ive binding approaches.  Second, an \emph{optional Numba JIT
path} with automatic AST-level loop parallelization distributes
user-written update logic across all available CPU cores without
requiring the developer to write explicit threading code.  Third, a
\emph{declarative render graph} with Python injection points allows
users to define, extend, and reorder rendering passes---including
custom post-processing and debug overlays---without leaving the
Python environment.

Benchmarks on a 24-core desktop with an RTX~5070\,Ti demonstrate
that the batch\,+\,JIT path sustains frame rates competitive with
Unity~6's IL2CPP-compiled C\# across rendering workloads up to
10\,000 draw calls, and achieves $6.9{\times}$ higher throughput on
pure-compute kernels at one million elements.  These comparisons are
subject to confounding factors---differing shading complexity, editor
maturity, and draw-call batching strategies---which we have discussed
explicitly.  Experiment~3 (pure compute) offers the cleanest
scripting-throughput comparison and shows the largest advantage for
the JIT path.  We have further identified the Python--C++ communication
boundary as the principal remaining bottleneck and outlined a lock-free
command-ring architecture that preliminary micro-benchmarks suggest can
lower crossing cost by at least an order of magnitude.

Infernux is MIT-licensed and publicly available at
\url{https://chenlizheme.github.io/Infernux/}.  We believe that
lowering the entry barrier to engine-level development through
Python's ecosystem---while maintaining the frame-rate expectations
of real-time applications---opens productive avenues for rapid
prototyping, educational use, and integration with the scientific
Python stack.

\section*{Acknowledgment}

This work was conducted at the Shenzhen International Graduate School,
Tsinghua University.

\bibliographystyle{IEEEtran}
\bibliography{references}

\end{document}